\def\aa{{ A\&A}}
\def\aas{{ A\&AS}}

\def\aj{{AJ}}

\def\al{$\alpha$}
\def\bet{$\beta$}

\def\annrev{{ARA\&A}} 

\def\apj{{ApJ}}

\def\apjs{{ApJS}}

\def\asec{$^{\prime\prime}$}

\def\cc{cm$^{-3}$}

\def\deg{$^{\circ}$}

\def\e#1{$\times$10$^{#1}$}
\def\etal{{et al. }}

\def\kms{km\thinspaces$^{-1}$}
\def\lamb{$\lambda$}
\def\lum{ergs s$^{-1}$}

\def\percm2{cm$^{-2}$}

\def\solmass{M$_\odot$}

\def\gax    {${_>\atop^{\sim}}$ }

\def\kms    {~km~s$^{-1}$}
\def\refindent{\par\noindent\parskip=2pt\hangindent=3pc\hangafter=1 }

\def\>Z{$>$}
\def\<{$<$}

\def\simlt{\lower.5ex\hbox{$\; \buildrel < \over \sim \;$}}
\def\simgt{\lower.5ex\hbox{$\; \buildrel > \over \sim \;$}}
\def\sqr#1#2{{\vcenter{\hrule height.#2pt
      \hbox{\vrule width.#2pt height#1pt \kern#1pt
         \vrule width.#2pt}
      \hrule height.#2pt}}}

\def\4363{[\ion{O}{3}]}

\def\oiii{[\ion{O}{3}]}

\def\oi{[\ion{O}{1}]}

\def\nii{[\ion{N}{2}]}

\def\sii{[\ion{S}{2}]}

\def\hii{\ion{H}{2}}

\documentstyle[11pt,aaspp4]{article}
\received{}
\accepted{}

\slugcomment{To appear in {\it The Astrophysical Journal}}

\begin{document}

\title{A Search for ``Dwarf'' Seyfert Nuclei. V. Demographics of Nuclear 
Activity in Nearby Galaxies}

\author{Luis C. Ho}
\affil{Department of Astronomy, University of California, Berkeley, CA 
94720-3411}

\and

\affil{Harvard-Smithsonian Center for Astrophysics, 60 Garden St., Cambridge,
MA 02138\footnote{Present address.}}
 
\author{Alexei V. Filippenko}
\affil{Department of Astronomy, University of California, Berkeley, CA
94720-3411}

\and

\author{Wallace L. W. Sargent}
\affil{Palomar Observatory, 105-24 Caltech, Pasadena, CA 91125}
 
\begin{abstract}
We use the sample of emission-line nuclei derived from a recently 
completed optical spectroscopic survey of nearby galaxies to quantify the 
incidence of local ($z\,\approx\,0$) nuclear activity.
Particular attention is paid to obtaining accurate measurements of 
the emission lines and reliable spectral classifications.  The resulting data 
base contains the largest collection of star-forming nuclei and active galactic 
nuclei (AGNs) currently known for nearby galaxies.  It consists of 420 
emission-line nuclei detected from a nearly complete, magnitude-limited 
sample of 486 galaxies with $B_T\,\leq\,12.5$ mag and declination $>$ 0\deg;
the equivalent-width detection limit of the brightest emission line, usually 
H\al, is $\sim$0.25 \AA.

Consistent with previous studies, we find detectable amounts of ionized gas in 
the central few hundred parsecs of most (86\%) galaxies; emission lines are 
present in essentially every spiral galaxy and in a large fraction of 
ellipticals and lenticulars.  Based on their 
narrow-line spectra, half of the objects can be classified as \hii\ or 
star-forming nuclei and the other half as some form of AGN, of which we 
distinguish three classes --- Seyfert nuclei, low-ionization nuclear 
emission-line regions (LINERs), and transition objects which we assume to 
be composite LINER/\hii-nucleus systems.  The population of AGNs consequently 
is very large; approximately 43\% of the galaxies in our survey can be 
regarded as ``active,'' although, for a number of reasons, this fraction is 
still rather uncertain.  Most of the objects have much lower 
luminosities than AGNs commonly studied; the median luminosity of the narrow 
H\al\ line, after correcting for extinction, is only 2\e{39} \lum.  Our sample 
therefore occupies the extreme faint end of the AGN luminosity function.  

We detect signatures of a broad-line region, as revealed by visible broad 
H\al\ emission, in $\sim$20\% of the AGN sample.  Seyfert nuclei, both 
type 1 and type 2, reside in $\sim$10\% of all galaxies.  LINERs 
make up the bulk (1/2--3/4) of the AGN population and a significant 
fraction (1/5--1/3) of all galaxies.  A nonnegligible subset of LINERs 
emit broad H\al\ emission, furnishing direct evidence that at least some 
LINERs are indeed related physically to the AGN phenomenon.

The dominant ionization mechanism of the nuclear emission depends strongly on 
the morphological type and luminosity of the host galaxy.  AGNs are found 
predominantly in luminous, early-type (E to Sbc) galaxies, while \hii\ 
nuclei prefer less luminous, late-type (Sbc and later) systems.  The various
AGN subclasses have broadly similar host galaxies.

\end{abstract}


\section{Introduction}

Emission-line spectroscopy of the central regions of galaxies can yield 
information often inaccessible through other observational techniques.  Optical
emission lines in particular trace the warm, ionized component of the 
interstellar medium.  In addition to providing information on the nebular 
conditions and kinematics of the line-emitting material, the emission 
lines, as reprocessed radiation, can potentially probe the physical 
mechanism responsible for the ionization of the gas.  The presence of 
optical and ultraviolet emission lines in galaxy nuclei is often taken to 
be a sign of nuclear ``activity,'' and spectroscopic surveys, especially 
at optical wavelengths, have become a widely practiced means of gathering 
large samples of emission-line nuclei for a variety of statistical studies.

One particular application has been to investigate the nature of the 
line emission in the central regions of nearby galaxies.  Over 
the last two decades,  a number of spectroscopic surveys of nearby 
galaxies have been conducted for this purpose (Heckman, Balick, \& Crane 1980; 
Heckman 1980b; Stauffer 1982a, b; Keel 1983a, b; Phillips \etal 1986; V\'eron 
\& V\'eron-Cetty 1986; V\'eron-Cetty \& V\'eron 1986).  One of the principal 
results of these studies is the realization that the incidence of nuclear 
activity, possibly nonstellar in origin, appears to be very high.  Heckman 
(1980b) identified low-ionization nuclear emission-line regions (LINERs) as 
major constituents of the extragalactic population, particularly among 
early-type galaxies.  The optical emission-line spectra of LINERs broadly 
resemble those of traditional active galactic nuclei (AGNs) such as 
Seyfert nuclei, but they have characteristically lower ionization levels.  The 
physical nature of LINERs has been the subject of considerable debate 
(see Filippenko 1996 for a recent review), but one viable interpretation is 
that they are simply another manifestation of the AGN phenomenon.  If true,
LINERs would heavily populate the faint end of the local luminosity function 
of AGNs, with consequences for a range of astrophysical issues.  In this 
paper we assume that LINERs are indeed genuine AGNs.

Not all emission-line nuclei require an exotic source of ionizing radiation.
A sizable fraction of the objects have spectra similar to those of giant 
extragalactic \hii\ regions, and their primary ionization mechanism must be 
photoionization by ultraviolet radiation from young, massive stars.  This 
population offers insights into the process of star formation in an 
environment that is likely to be substantially different from that of galactic 
disks.

The above-mentioned surveys, while tremendously valuable in establishing 
the qualitative patterns of nuclear activity among nearby galaxies, suffer 
from several shortcomings that make quantitative applications uncertain. 
At optical wavelengths the nuclear component of a typical nearby galaxy is 
generally much weaker than the stellar background of its bulge.  Thus, in 
addition to having small fluxes and sometimes being blended together, 
the emission lines are diluted by stellar absorption lines, necessitating 
careful removal of the starlight contamination for accurate measurements.  
As discussed by Ho (1996), this crucial step in the analysis was not always 
treated adequately in many of the older studies.  

We recently completed an extensive spectroscopic study of the nuclear 
regions of nearly 500 bright, northern galaxies.  This survey contains the 
largest published data base of homogeneous and high-quality optical spectra of 
nearby galaxies; it represents a significant improvement, both in 
sample size and in sensitivity, compared with previous studies of its kind.  
We have invested substantial effort to correct the spectra for starlight 
contamination in a consistent and objective fashion.  In addition to being 
able to detect much fainter emission lines than has been possible in 
the past, we believe that our emission-line measurements are quantitatively 
much more reliable.  This distinction directly impacts the accuracy of the 
spectral classification, with ramifications for all ensuing analyses that make 
use of the statistics of the various classes of emission-line nuclei. 

The purpose of this paper is to summarize the demographics of emission-line 
nuclei in light of these new data.  Specifically, we report on the detection 
rates of star-forming nuclei and of various subclasses of AGNs, and we 
examine the dependence of their detection rates and number distributions 
on the morphological type and luminosity of the host galaxies.  The 
likely influence of selection biases and sample incompleteness are discussed.
Some general statistical properties of the sample are additionally noted.

\section{The Palomar Survey}

The analysis in this paper is based on a magnitude-limited survey of 486 
northern galaxies.  The sample is defined to be all galaxies listed in the 
Revised Shapley-Ames Catalog of Bright Galaxies (RSA; Sandage \& Tammann 1981) 
with $\delta\,>$ 0\deg\ and $B_T\,\leq$ 12.5 mag, with a few minor alterations 
as described by Ho, Filippenko, \& Sargent (1995, hereafter Paper II).  The 
data base consists of high-quality optical spectra of moderate resolution 
(100--200 \kms) acquired with the Hale 5~m telescope at Palomar Observatory 
(Filippenko \& Sargent 1985, hereafter Paper I).  The selection criteria of 
the survey ensure that the sample is a fair representation of the local 
($z\,\approx\,0$) galaxy population, at least for high-surface brightness 
systems, and the proximity of the objects enables fairly good spatial 
resolution to be achieved.  We employed a long slit of width 2\asec\ and 
adopted an extraction width of 4\asec, which projects to an aperture with 
linear dimensions $\sim200\,\times\,400$ pc$^2$ for the typical distances of 
the sample galaxies (18 Mpc; Table 1).\footnote{We adopt $H_0$ = 75 \kms\ 
Mpc$^{-1}$ in this series of papers.}  Paper II presents 
the spectral atlas of the survey and discusses the observational parameters and 
data reduction; Paper III (Ho, Filippenko, \& Sargent 1997a) gives the line 
measurements, object classifications, and details of our treatment of 
starlight subtraction; Paper IV (Ho \etal 1997e) highlights the nuclei 
showing broad H\al\ emission; and Paper VI (Ho, Filippenko, \& Sargent 1997b) 
provides a comparative analysis of the various AGN subclasses.  Additional 
papers in this series (Ho, Filippenko, \& Sargent 1997c, d) analyze the 
subsamples of star-forming nuclei and barred galaxies.  All quantities
used in this study are drawn from Paper III.

The classification system used throughout our survey parallels closely the 
methodology of Veilleux \& Osterbrock (1987).  As explained in Paper III, this 
system adopts a set of spectroscopic criteria that depends entirely on the 
line-intensity ratios of several prominent, narrow, optical emission lines.  
We distinguish four subclasses of emission-line nuclei: \hii\ nuclei, Seyfert 
nuclei, LINERs, and transition objects.  \hii\ nuclei have spectra closely 
resembling those of \hii\ regions and therefore are assumed to be powered 
through photoionization by young, massive stars.  The other three groups 
represent variants of AGNs.  The composite characteristics of the spectra of 
transition objects suggest that they are LINER nuclei contaminated by emission 
from neighboring \hii\ regions (Ho, Filippenko, \& Sargent 1993; Ho 1996); 
this hypothesis is explored further in Paper VI, where we demonstrate that 
transition objects and regular LINERs have strikingly similar global and 
nuclear properties, suggesting that they share a common physical origin.  
However, it should be borne in mind that the available data cannot yet
unambiguously exclude alternative explanations that do not invoke nonstellar 
processes.  In the following discussion, we will explicitly 
assume that transition objects indeed contain LINER nuclei, and hence 
that they are AGNs, although we will point out how the results would be 
affected if this assumption were to be relaxed.  When there is a need to 
distinguish regular LINERs from composite sources, we 
will refer to the former as ``pure LINERs'' and to the combined sample of 
pure LINERs and transition objects as ``all LINERs.''  Finally, all three 
classes of AGNs can host a broad-line region, as evidenced by the presence 
of broad H\al\ emission (Paper IV).  Following the convention of Papers III 
and IV, we extend the ``type 1'' and ``type 2'' designations of Seyfert 
galaxies (Khachikian \& Weedman 1974) to include LINERs and transition 
objects.  

\section{Detection Rates of Emission-Line Nuclei}

The incidence of emission-line nuclei is very high in our sample (Table 
2{\it a} and top panel of Fig. 1{\it a}).  Integrated over all Hubble types, 
86\% of the nuclei have emission lines down to an equivalent-width detection 
limit of $\sim$0.25 \AA\ (3 $\sigma$).  The detection rate among spirals alone 
is even higher: essentially all (98\%) of the galaxies classified as S0/a and 
later have emission-line nuclei, to be compared with 54\% for the ellipticals 
and 64\% for the lenticulars.  Among the group of 66 galaxies with pure 
absorption-line spectra, only 8 are not classified as ellipticals or 
lenticulars.  In a sample of disk systems with Hubble types ranging from S0/a 
to Scd, but with a brighter limiting magnitude ($B_T\,\leq\,12.0$ mag), Keel 
(1983a) also found that essentially every object has detectable emission lines 
within an 8\asec\ circular aperture.  Because the sensitivity of the Palomar 
survey is much higher than that of Keel, we are able to achieve a comparably 
high detection rate for this range of morphological types, even though our 
effective aperture (8 arcsec$^2$) is six times smaller and our survey limit 
fainter.  The Hubble type distributions of the surveys of Heckman \etal (1980) 
and V\'eron-Cetty \& V\'eron (1986) more closely match that of the present
sample, and, in these, the detection rate was only $\sim$60\%--65\%.  The near 
ubiquity of emission lines in the nuclear spectra implies that the central 
200--400 pc of most galaxies, including those of early type, contain detectable 
amounts of warm ($\sim 10^4$ K) ionized gas.  The typical H\al\ luminosity of 
$\sim$1\e{39} \lum\ (\S\ 6) and electron density of 200 \cc\ (Paper VI; 
Ho \etal 1997c) translate to an ionized hydrogen mass of $\sim$2\e{4} 
\solmass.

The two categories of nuclear activity (stellar and nonstellar) occur with 
nearly equal frequency among galaxy nuclei.  Approximately half of the 
emission-line objects (42\% of all galaxies) are classified as \hii\ nuclei, 
and the other half belongs to the AGN group (43\% of all galaxies); the 
proportion becomes 56\% \hii\ nuclei and 30\% AGNs if we reassign 
transition objects to the former group.  While the 
incidence of both varieties of nuclear activity is widespread among galaxies 
of all morphologies, each depends strongly and differently on the Hubble type 
of the host (middle and bottom panels of Fig. 1{\it a}).  \hii\ nuclei clearly 
prefer late-type hosts, whereas AGNs prefer early-type hosts.  Some overlap 
occurs between the two distributions, but they segregate roughly at a Hubble 
type of Sbc: 82\% of galaxies later than Sbc have \hii\ nuclei, and 60\% of 
galaxies earlier than Sbc have AGNs.  Surprisingly, not a single 
elliptical galaxy in our sample shows detectable nuclear emission attributable 
to star formation, in stark contrast to the substantial fraction of AGNs that 
contribute to this morphological bin ($\sim$50\%).  This is consistent with 
the survey of early-type (E and S0) galaxies of Phillips \etal (1986); the few 
objects they identified as having \hii\ nuclei are all classified S0 (two are 
E-S0).  Signatures of nonstellar ionization, on the other hand, do exist in a 
minority of late-type hosts; roughly 15\% of the Sc, Sd, and Sm galaxies, and 
40\% of the amorphous systems (I0) contain AGNs (but there are only 
five amorphous galaxies in our sample, so the latter statistic should be 
treated with caution).

\section{Subclasses of AGNs}

All three subclasses of AGNs show similar detection rates as a function of 
Hubble type (Fig. 1{\it b}).  The most conspicuous differences are that (a) 
pure LINERs, compared to Seyferts, are seen in a higher fraction of ellipticals 
and, (b) among all LINERs, the transition group is detected more frequently in 
galaxies of somewhat later Hubble types.  Approximately 10\% of the survey 
sample contain Seyfert nuclei; this doubles the figures estimated in previous 
studies (Stauffer 1982b; Keel 1983b; Phillips, Charles, \& Baldwin 1983; 
Maiolino \& Rieke 1995).  Note that the Seyfert nuclei in 
our sample do not exclusively reside in spirals, as is usually thought (e.g., 
Adams 1977; Weedman 1977).  Pure LINERs are present in $\sim$19\% of all 
galaxies, and transition objects, which by assumption also contain a LINER 
component, account for another $\sim$13\%. Thus, LINERs are major constituents 
of the galaxy population --- they reside in 
1/3 of all galaxies brighter than $B_T$ = 12.5 mag.  Because of the 
strong preference for early-type hosts, the detection rate approaches 50\% for 
galaxies of types E--Sbc. If all LINERs can be regarded as genuine AGNs, they 
make up the bulk of the AGN population (75\%) in the luminosity range probed 
by our survey, outnumbering Seyferts two to one. 

A sizable fraction of the AGN sample ($\sim$20\%) shows broad H\al\ emission, 
presumably arising from the conventional broad-line region (Paper IV).  
The broad emission is generally very weak and difficult to measure; 
consequently, most of the objects identified in our survey have previously 
been unrecognized. Of the 46 detections reported in Paper IV, only 22 formally 
have a Seyfert classification, and the remaining 24 fall in the LINER group 
(22 LINERs and 2 transition objects).  We proposed in Paper IV that the 
``type 1/type 2'' designations, traditionally used to distinguish between 
Seyfert nuclei with and without a visible broad-line region, respectively, be 
extended to include LINERs and transition objects.  The number ratio of type 2 
to type 1 Seyferts in our survey is 1.4 to 1; the corresponding ratio for 
pure LINERs is 3.3 to 1, and for all LINERs (including transition 
objects) it is 5.6 to 1.  As discussed in Paper IV, we suspect that 
selection effects severely hamper the detection of broad H\al\ in transition 
objects, thereby leading to an apparently low incidence of type 1 objects 
in this group.  It is possible that the true frequency of type 1 transition 
objects is as high as that of type 1 pure LINERs.

\section{Trends with Galaxy Morphological Type and Integrated Luminosity}

The distribution of morphological types in Figure 2{\it a} illustrates that 
\hii\ nuclei reside most frequently in Sc galaxies [median T = 5.0, where T is 
the numerical Hubble type index as defined by de Vaucouleurs (1959, 1963)], and 
most AGNs cluster toward early-type disks systems (S0--Sbc; median T = 1.0, 
corresponding to Sa).  The three AGN subclasses once again show very similar 
distributions of host galaxy types (Fig. 2{\it b}).  LINERs and Seyferts have 
virtually indistinguishable host galaxy types (aside from a higher proportion 
of ellipticals among LINERs), an important clue to the physical nature of 
LINERs (Paper VI).

The frequency of bars among the emission-line objects is identical to that of 
the entire sample of disk systems in the survey (56\%; Paper III), since most 
of the absorption-line objects are ellipticals.  The bar fraction of the 
\hii\ nuclei hosts (62\%) does not differ appreciably from that of the 
AGN hosts (49\%), which itself remains constant among the AGN subclasses.  
However, as discussed more fully by Ho \etal (1997d), the presence of a 
bar does enhance the probability and intensity of nuclear star formation
in spiral galaxies.  Such an effect is not seen among the AGN hosts.

Because early-type galaxies on average tend to be more luminous compared 
to late-type galaxies (e.g., Roberts \& Haynes 1994), the trends perceived 
with Hubble type translate into similar patterns in total galaxy 
luminosity.  The distributions of absolute blue magnitudes, corrected for 
internal extinction ($M_{B_T}^0$; Paper III), are shown in Figure 3{\it a}.  
The hosts of \hii\ nuclei clearly have lower luminosities than the 
hosts of AGNs, being fainter than the latter by $\sim$0.5 mag in their median 
$M_{B_T}^0$ (--20.01 mag versus --20.46 mag).  The cumulative distributions of 
the two samples are significantly different according to the 
Kolmogorov-Smirnov test (Press \etal 1986); the probability that the two 
distributions are drawn from the same population ($P_{\rm KS}$) is 5.7\e{-4}.  

Interestingly, the objects lacking emission-line nuclei are 
noticeably less luminous (median $M_{B_T}^0$ = --19.56 mag) than those 
containing either \hii\ nuclei or AGNs.  Since almost all of the 
absorption-line objects are ellipticals and lenticulars, a fair comparison 
sample should be restricted to have the same range of Hubble types.  The 
sample of emission-line objects classified as E and S0 has a significantly 
higher median luminosity than the absorption-line objects 
($\Delta M_{B_T}^0$ = 0.66 mag; $P_{\rm KS}$ = 0.0046).  The difference 
remains even after excluding the two extremely low-luminosity dwarf elliptical 
galaxies (NGC 147 and NGC 205) from the absorption-line sample.  The 
availability of gas, a necessary condition for the generation of emission 
lines, somehow appears to depend on the total luminosity, and presumably mass, 
of the galaxy.  We are tempted to speculate that the supply of gas is linked 
to the mass of the stellar component, possibly via mass loss from evolved stars.

The homogeneity noted in the morphologies of the host galaxies of the AGN 
subclasses becomes even more striking when we examine their absolute 
magnitudes (Fig. 3{\it b}).  Apart from a slight excess of low-luminosity 
objects among the transition objects, all three AGN groups have similar 
distributions of absolute magnitudes; any minor differences among them have no 
formal statistical significance.  It is again noteworthy that LINERs and 
Seyferts both peak at $M_{B_T}^0\,\approx$ --20.5 mag, about 0.4 mag 
brighter than $M_{B_T}^*$, the typical absolute magnitude of the field-galaxy 
luminosity function (e.g., Kirshner, Oemler, \& Schechter 1979, after adjusting
to our adopted $H_0$ = 75 \kms\ Mpc$^{-1}$).

\section{Strengths of the Emission Lines}

Although the emission-line properties of the AGN and \hii-nuclei samples 
are discussed in separate publications (Paper VI; Ho \etal 1997c), here we will 
comment briefly on the strengths of the emission lines.  In general, the line 
emission of the objects in the Palomar survey is quite feeble.  A wide range 
of equivalent widths is found, but the median value for the H\al\ line 
is only 5 \AA\ (Fig. 4).  \hii\ nuclei have significantly higher emission-line 
equivalent widths than AGNs [median EW(H\al) = 18 \AA\ versus 2 \AA].  
The marked contrast in equivalent widths between the two classes of 
nuclei arises not because of intrinsic luminosity differences, but 
rather because of the great disparity between the nuclear (stellar) continuum 
strengths of the two types of host galaxies.  Late-type galaxies, the 
preferred hosts of \hii\ nuclei, have fainter, smaller bulges than early-type 
galaxies, the dominant hosts of AGNs.  The extinction-corrected H\al\ 
luminosities of \hii\ nuclei are in fact larger than those of the AGN 
sample (Fig. 5), but the difference is only a factor of two [median 
L(H\al) = 1.8\e{39} \lum\ versus  8.5\e{38} \lum], whereas the difference 
in the equivalent widths of the two groups amounts to a factor of nine.
The variation of the emission-line equivalent width with galaxy type and 
luminosity is illustrated in Figure 6, where the steady rise of the relative
line strength toward galaxies with later Hubble types and lower luminosities 
is quite apparent.  By contrast, the line luminosity actually decreases 
in late-type and low-luminosity galaxies (Fig. 7).

The majority of the \hii\ nuclei in our survey 
are experiencing only modest levels of current star formation.  Indeed, the 
typical H\al\ luminosity does not greatly exceed that of many individual 
giant \hii\ regions, and the inferred current star-formation rates certainly 
are not unusual.  Thus, we have resisted calling these objects ``starburst'' 
nuclei like those of Balzano (1983). Similarly, the AGNs considered here have 
unspectacular luminosities when compared to traditionally studied Seyferts 
such as those selected from the Markarian survey.  The Seyferts in the 
compilation of Dahari \& De Robertis (1988), for instance, have typical line 
luminosities ranging from two to three orders of magnitude larger than those 
in our survey.  Our sample, therefore, contains mainly {\it low-luminosity} 
or ``dwarf''  AGNs.

\section{Completeness and Selection Effects}

The completeness of the overall Palomar survey is very close to that of a 
sample limited to $B_T\,\leq$ 12.5 mag in the RSA, from which our sample was 
drawn.  A discussion of the completeness of the RSA can be found in 
Sandage, Tammann, \& Yahil (1979).  Here we wish to consider the completeness 
of the different types of emission-line objects with respect to the parent 
population.  A simple way to examine this issue is to compare the 
distribution of apparent magnitudes of the different subsamples with that of 
the parent sample.  From Figure 8{\it a}, it is evident that the parent sample 
and the sample of \hii\ nuclei hosts have very similar distributions of $B_T$ 
($P_{\rm KS}$ = 0.13), indicating that the latter is not incomplete relative 
to the former.  The AGN sample, on the other hand, has a marginally 
brighter $B_T$ distribution than the parent sample ($P_{\rm KS}$ = 0.043).
This arises not because the AGN sample has on average smaller distances 
(see below), but rather because it is comprised mainly of more luminous, 
early-type galaxies (\S\ 5).  The AGN sample, therefore, suffers from some 
incompleteness with respect to the parent sample, although the effect seems 
to be slight.  On closer inspection, it appears that most of the difference 
can be attributed to the group of pure LINERs alone (Fig. 8{\it b}).

To assess how the detection rates presented thus far would be modified 
by the relative incompleteness of the different types of nuclei, we chose 
subsamples with various brighter apparent magnitude 
limits from the parent sample, recomputed the detection rate of each 
group of emission-line object, and calculated the Kolmogorov-Smirnov 
statistic to gauge the change in completeness levels.  This experiment
showed that all the emission-line groups become complete with respect to 
the new parent subsamples at $B_T\,\approx$ 12.2--12.3 mag --- that is 
to say, there is no statistically significant difference ($P_{\rm KS}\,>\,
0.1$) in the relative distributions of $B_T$.  The detection rates of the 
various types of emission-line nuclei at this new limiting magnitude, 
however, hardly change from those found using the original limiting magnitude.  
The detection rate for AGNs, and similarly for LINERs and transition objects, 
increases by $\sim$10\%, that of \hii\ nuclei decreases by the same amount, 
and that of Seyferts remains essentially unaltered.  

Because our data were acquired using a slit of fixed angular size, the physical
dimensions of the projected aperture scale linearly with the distance 
of the object, and distance-dependent selection biases in principle can 
affect our measurements.  Specifically, with regard to the detection rates 
under consideration, two kinds of selection effects can occur.  First, 
the detection of any line emission, regardless of its character, obviously 
depends on the angular dimension of the emission compared to the aperture size.
For a given (low) surface brightness, the emitting material can be 
undetectable if it is very extended compared to the aperture, for example if 
the galaxy is exceptionally nearby.  The object would then be 
considered to lack emission lines, even though it would have been recognized 
as an emission-line nucleus had it been placed at a distance more typical 
of the rest of the sample.  An interesting example is M31.  We 
failed to detect any line emission in the spectrum of its nucleus because of 
its proximity (0.75 Mpc), and in Paper III we classified it as an 
absorption-line nucleus.  However, line emission of low surface brightness, 
extending over scales of several hundred parsecs, {\it does} exist in the 
circumnuclear regions of M31 (Rubin \& Ford 1971).  Moreover, Ciardullo \etal 
(1988) have shown that the spectrum of the gas shows enhanced 
\nii\ \lamb\lamb 6548, 6583 and \sii\ \lamb\lamb 6716, 6731 emission, as is 
typical of most AGNs (see Paper III).  Heckman (1996) recently concluded that 
the spectrum is that of a LINER.  The effect of distance, however, has a 
negligible bearing on our results because the absorption-line sample does not 
have a smaller median distance than the emission-line sample, and because 
the detection rate of emission-line objects is already so high (86\%) that 
there is not much room for error.  

Perhaps more worrisome is the accuracy of the {\it relative} detection rates 
among the emission-line objects.  The integrated spectrum of the central 
region of a distant galaxy has a higher likelihood of being contaminated by 
circumnuclear \hii\ regions than a nearby one, thereby potentially biasing the 
AGN detection rate toward lower values among distant galaxies.  However, the 
distributions of distances show no gross differences for the various 
subclasses (Figs. 9{\it a} and 9{\it b}). The \hii\ nuclei are on average 
closer than the AGNs (median distance 17.1 Mpc versus 20.6 Mpc; $P_{\rm KS}$ 
= 2.6\e{-4}) as a result of having lower luminosity hosts, and LINERs (both 
including and excluding transition objects) are marginally more distant than 
Seyferts (by 1--2 Mpc).  In any case, it seems unlikely that such small 
differences in distances can lead to significant misclassifications in the 
mean.  We reached a similar conclusion in Paper III based on analysis of the 
variation of the \nii\ \lamb 6583/H\al\ ratio with distance.  Any individual 
object, of course, can certainly still be affected.  We mentioned the case 
of M31 above.  Rubin \& Ford (1986) discussed a similar situation for the 
nucleus of M33.

Finally, in Figure 10 we examine the inclination angles ($i$) of the disk 
systems.  The distribution of cosine $i$ should be flat for an unbiased 
sample with random orientations.  As discussed in Paper III, there is a 
deficit of edge-on systems ($i$ \gax 70\deg) in the parent sample, and this 
behavior is characteristic of magnitude-limited samples.  Again, what is of 
interest here is to see if there are any differences between the 
total sample and each of the separate groups, as well as among the groups. 
None of the subsamples, with the exception of the Seyferts, 
show statistically different distributions of cosine $i$ compared to the total 
sample. Seyferts do show a marginally significant deficit of edge-on systems 
($P_{\rm KS}$ = 0.047), and they are also somewhat more face-on than the 
combined LINER sample ($P_{\rm KS}$ = 0.060).  Another subtle difference 
is that transition objects tend to be more edge-on than pure LINERs 
($P_{\rm KS}$ = 0.064).  Although these differences are not large, 
they are of relevance in understanding the physical distinctions between the 
AGN subclasses, and we will reconsider them in Paper VI.  But, for 
now, we simply note that selection biases due to inclination effects 
do not appear to be serious.

In summary, we consider the detection rates reported in Table 2 
and Figure 1, in both absolute and relative numbers, to be largely 
uncorrupted by incompleteness introduced either by the magnitude limit of the 
survey or by selection effects due to distance or inclination angle.

\section{Comparison with Previous Studies}

Many of the findings from the Palomar survey presented here are qualitatively 
similar to results from the older surveys cited in \S\ 1.  It has long been 
recognized that the incidence of nuclear activity, especially as evidenced by 
the LINER phenomenon, is widespread in the nearby galaxy population.  Although 
it is still unclear whether all LINERs can be unequivocally associated with 
AGNs (Paper VI), the general consensus has been that these objects must be 
related to some form of activity substantially different from ``normal'' star 
formation.  Furthermore, it has occasionally been pointed out that 
galaxies hosting \hii\ nuclei have quite different morphological types 
than those containing Seyfert or LINER nuclei (e.g., Heckman 1980a; Keel 
1983a; Terlevich, Melnick, \& Moles 1987; Pogge 1989).  Indeed, Burbidge \& 
Burbidge (1962) drew attention to the fact that some galaxies show abnormal 
strengths of \nii\ \lamb 6583 compared to H\al, and they noted that such 
galaxies tend to be of early type.  

As discussed in \S\ 1, the Palomar survey has greater sensitivity to weak 
emission lines than previous surveys of this kind.  From a statistical 
perspective, it also contains a larger number of galaxies as well as a 
wider range of morphological types [see the summary of old surveys presented 
in Table 1 of Ho (1996)].  More importantly, however, we believe our 
emission-line measurements, and hence all subsequent derivations, to be
{\it quantitatively} much more reliable.  Our spectral classification, 
in particular, should be considerably more secure.  The increased accuracy 
largely stems from our treatment of starlight correction.  As an 
example of the immediate benefits to be gained, note that the previous studies 
rarely were able to detect the weak, but diagnostically important, 
\oi\ \lamb 6300 line.  Our ability not only to detect, but to measure \oi\ in 
a significant fraction of our emission-line objects (81\%) has led us 
recognize the class of sources we call transition objects.  Having access to 
a wider wavelength range, particularly in the blue, further allows us to 
better specify the classification.  The surveys of Keel (1983a, b) and 
Phillips \etal (1986), for instance, did not include the H\bet\ and \oiii\ 
\lamb\lamb 4959, 5007 lines, so they had no information on the excitation of 
their emission-line objects, and therefore no way to distinguish between 
Seyferts and LINERs.  Perhaps the most dramatic improvement, however, can be 
seen in the high detection rate of broad H\al\ emission in our survey (Paper 
IV).  This has resulted in a robust determination of the relative fraction of 
type 1 and type 2 AGNs, and it has shown, for the first time, that a 
significant fraction of LINERs contain a broad-line region, a finding that has 
important consequences for the longstanding debate on the physical origin of 
this class of objects (Paper VI).

As is well known, existing AGN samples suffer from various degrees of biases 
and incompleteness [see, e.g., discussion in Huchra \& Burg (1992)].  The 
incompleteness is most severe for low-luminosity sources.  The sample of 
Seyfert nuclei spectroscopically selected from the CfA redshift survey 
(Huchra \& Burg 1992) is widely regarded as perhaps the least biased available 
set.  Yet, even this sample misses many of the weak sources included in the 
Palomar list, and, as recognized by Huchra \& Burg (1992), the CfA sample is 
very incomplete in its census of LINERs.  Maiolino \& Rieke (1995) improved 
the situation by tallying the Seyfert content in the RSA ($B_T\,<\,13.4$ mag) 
based on spectral classifications taken from the literature.  They deduced a 
lower limit of 5\% to the frequency of Seyfert nuclei in nearby galaxies, but, 
based on completeness considerations, they argue that the true frequency could 
be as high as 16\%.  Their lower limit, while consistent with our results, is 
too low by a factor of $\sim$2, and their estimate of the true frequency 
appears to be somewhat high.  Since Maiolino \& Rieke based their study on 
published material, it is not surprising that they, too, missed many of the 
Seyferts recovered in our survey.  Indeed, a significant fraction of the 
published classifications they used were drawn from the very studies which we 
evaluated relative to the Palomar survey in \S\ 1.  Only half of the 52 
Seyferts in the Palomar sample appear in the tabulation of Maiolino \& Rieke.  

\section{Concluding Summary}

A large sample of emission-line nuclei has been identified in a recently
completed optical spectroscopic survey of nearby galaxies, allowing several
statistical properties of the host galaxies and of the line-emitting
regions to be examined reliably for the first time.  As a consequence of
the many detections and some revised classifications, the detailed
demographics of emission-line nuclei have been updated from those given in
older surveys.  Table 1 gives a synopsis of their main characteristics, and 
Table 2 summarizes the detection rates of the different object classes.
This paper reports the detection rate of line emission in the central regions 
of galaxies, the incidence of different classes of emission-line nuclei, and 
their dependence on the morphological type and luminosity of the 
host galaxy type.  The main conclusions of this paper, which are based on 
420 emission-line nuclei selected from a magnitude-limited ($B_T\,\leq\,12.5$ 
mag) sample of 486 northern ($\delta\,>$ 0\deg) galaxies, are as follows.

(1) Consistent with previous studies, the central few hundred parsecs of 
most (86\%) galaxies have detectable amounts of ionized gas as traced 
by optical emission lines.  The detection rate essentially reaches 100 
percent for spiral galaxies.

(2) The emission-line nuclei divide nearly equally in number between \hii\ 
nuclei and AGNs, where AGNs collectively refer to Seyfert nuclei, LINERs, 
and transition objects (composite LINER/\hii\ nuclei).  The AGN fraction in 
nearby galaxies is therefore very high, on the order of 43\%.  Unfortunately, 
the incidence of galaxies harboring a central massive black hole is still quite 
uncertain.  The AGN fraction could be considerably lower, for instance, if 
it turns out that many transition objects do not contain LINER nuclei, or 
if only a minority of LINERs are genuine AGNs.  We argue in Paper VI that 
this is unlikely to be the case.  On the other hand, very weak AGNs can be 
hidden by brighter nuclear \hii\ regions, or they may contain undetectably 
small amounts of ionized gas, and so at least some faint objects undoubtedly 
must have escaped notice.  Efforts to quantify these effects are in progress.

(3) Based on the relative intensities of the narrow emission lines, at least 
10\% of all galaxies in the present survey are classified as Seyfert nuclei 
(types 1 and 2).

(4) LINERs are found in 1/5 to 1/3 of all galaxies and, under the assumption 
that they are genuine AGNs, they constitute between 1/2 to 3/4 of the AGN 
population, depending on whether transition objects are excluded or included 
in the LINER group.  

(5) Broad-lined or ``type 1'' AGNs make up $\sim$20\% of the AGN population 
and $\sim$10\% of all galaxies.  Approximately half of the type 1 objects 
belong to the LINER category.

(6) The dominant excitation mechanism of the nuclear emission depends 
strongly on the Hubble type and integrated luminosity of the host galaxy.  AGNs 
reside mainly in early-type (E to Sbc) galaxies, while \hii\ nuclei prefer 
late-type (Sbc and later) systems.  AGN hosts are more luminous than 
non-AGN hosts because early-type galaxies tend to be more luminous than 
late-type galaxies.

(7) The subclasses of AGNs have broadly similar distributions of host 
galaxy morphological types and luminosities.  The only noticeable difference 
is that a higher proportion of pure LINERs is found in elliptical galaxies, 
while a higher fraction of transition objects tends to be in late-type hosts.

(8) The typical object has quite modest emission-line strengths, with H\al\
equivalent widths of only a few \AA, and luminosities (after correcting 
for extinction) of $\sim 10^{39}$ \lum.

(9) The detection rates of the various classes of emission-line objects are 
unlikely to be seriously incomplete or affected by selection biases due to 
distance or inclination.

\acknowledgments

We thank the referee, Tim Heckman, for helpful comments on the manuscript.
The research of L.~C.~H. is currently funded by a postdoctoral fellowship
from the Harvard-Smithsonian Center for Astrophysics.  Financial support for
this work was provided by NSF grants AST-8957063 and AST-9221365, as well as
by NASA grants AR-5291-93A and AR-5792-94A from the Space Telescope Science
Institute (operated by AURA, Inc., under NASA contract NAS5-26555).

\clearpage

\centerline{\bf{References}}
\medskip

\refindent 
Adams, T.~F. 1977, \apjs, 33, 19

\refindent 
Balzano, V.~A. 1983, \apj, 268, 602

\refindent
Burbidge, E.~M., \& Burbidge, G. 1962, \apj, 135, 694

\refindent
Ciardullo, R., Rubin, V.~C., Jacoby, G.~H., Ford, H.~C., \& Ford, W.~K., Jr. 
1988, \aj, 95, 438

\refindent
Dahari, O., \& De Robertis, M.~M. 1988, \apjs, 67, 249

\refindent
de Vaucouleurs, G. 1959, Handbuch der Physik, 53, 275

\refindent
de Vaucouleurs, G. 1963, \apjs, 8, 31

\refindent 
Filippenko, A.~V. 1996, in The Physics of LINERs in View of Recent 
Observations, ed.  M. Eracleous, et al. (San Francisco: ASP), 17

\refindent 
Filippenko, A.~V., \& Sargent, W.~L.~W. 1985, \apjs, 57, 503 (Paper I)
 
\refindent 
Heckman, T.~M. 1980a, \aa, 87, 142

\refindent 
Heckman, T.~M. 1980b, \aa, 87, 152
 
\refindent 
Heckman, T.~M. 1996, in The Physics of LINERs in View of Recent Observations, 
ed. M. Eracleous, et al., (San Francisco: ASP), 241

\refindent 
Heckman, T.~M., Balick, B., \& Crane, P.~C. 1980, \aas, 40, 295

\refindent 
Ho, L.~C. 1996, in The Physics of LINERs in View of Recent Observations, ed.
M. Eracleous, et al. (San Francisco: ASP), 103

\refindent 
Ho, L.~C., Filippenko, A.~V., \& Sargent, W.~L.~W. 1993, \apj, 417, 63
 
\refindent 
Ho, L.~C., Filippenko, A.~V., \& Sargent, W.~L.~W. 1995, \apjs, 98, 477 
(Paper II)

\refindent 
Ho, L.~C., Filippenko, A.~V., \& Sargent, W.~L.~W. 1997a, \apjs, in press
(Paper III)
 
\refindent 
Ho, L.~C., Filippenko, A.~V., \& Sargent, W.~L.~W. 1997b, in preparation
(Paper VI)

\refindent 
Ho, L.~C., Filippenko, A.~V., \& Sargent, W.~L.~W. 1997c, \apj, in press
 
\refindent 
Ho, L.~C., Filippenko, A.~V., \& Sargent, W.~L.~W. 1997d, \apj, in press

\refindent
Ho, L.~C., Filippenko, A.~V., Sargent, W.~L.~W., \& Peng, C.~Y. 1997e, \apjs, 
in press (Paper IV)

\refindent 
Huchra, J.~P., \& Burg, R. 1992, \apj, 393, 90

\refindent 
Keel, W.~C. 1983a, \apjs, 52, 229

\refindent 
Keel, W.~C. 1983b, \apj, 269, 466
 
\refindent 
Khachikian, E.~Ye., \& Weedman, D.~W. 1974, \apj, 192, 581

\refindent 
Kirshner, R.~P., Oemler, A., Jr., \& Schechter, P.~L. 1979, \aj, 84, 951

\refindent 
Maiolino, R., \& Rieke, G.~H. 1995, \apj, 454, 95

\refindent
Phillips, M.~M., Charles, P.~A., \& Baldwin, J.~A. 1983, \apj, 266, 485

\refindent 
Phillips, M.~M., Jenkins, C.~R., Dopita, M.~A., Sadler, E.~M., \&
Binette, L. 1986, \aj, 91, 1062
 
\refindent
Pogge, R.~W. 1989, \apjs, 71, 433

\refindent
Press, W.~H., Flannery, B.~P., Teukolsky, S.~A., \& Vetterling, W.~T. 1986,
Numerical Recipes (Cambridge: Cambridge Univ. Press)

\refindent 
Roberts, M.~S., \& Haynes, M.~P. 1994, \annrev, 32, 115

\refindent 
Rubin, V.~C., \& Ford, W.~K., Jr. 1971, \apj, 170, 25

\refindent 
Rubin, V.~C., \& Ford, W.~K., Jr. 1986, \apj, 305, L35

\refindent 
Sandage, A.~R., \& Tammann, G.~A. 1981, A Revised Shapley-Ames Catalog of
Bright Galaxies (Washington, DC: Carnegie Institute of Washington) (RSA)

\refindent 
Sandage, A.~R., Tammann, G.~A., \& Yahil, A. 1979, \apj, 232, 352

\refindent 
Stauffer, J.~R. 1982a, \apjs, 50, 517

\refindent 
Stauffer, J.~R. 1982b, \apj, 262, 66

\refindent 
Terlevich, R., Melnick, J., \& Moles, M. 1987, in Observational Evidence of
Activity in Galaxies, ed. E.~Ye. Khachikian, K.~J. Fricke, \& J. Melnick
(Dordrecht: Reidel), 499

\refindent 
Veilleux, S., \& Osterbrock, D.~E. 1987, \apjs, 63, 295

\refindent 
V\'eron, P., \& V\'eron-Cetty, M.-P. 1986, \aa, 161, 145

\refindent 
V\'eron-Cetty, M.-P., \& V\'eron, P. 1986, \aas, 66, 335

\refindent 
Weedman, D.~W. 1977, \annrev, 15, 69

%

\clearpage

\centerline{\bf{Figure Captions}}
\medskip

Fig. 1. --- 
Detection rate as a function of Hubble type of ({\it a}) all emission-line 
nuclei, \hii\ nuclei, and AGNs, and ({\it b}) the different classes of AGNs.  
The bins along the abscissa have the following meanings: ``E'' = E, 
``S0'' = S0, ``Sa'' = S0/a--Sab, ``Sb'' = Sb--Sbc, ``Sc'' = Sc--Scd, ``Sd'' 
= Sd--Sdm, ``Sm'' = Sm--Im, ``I0'' = I0, and ``Pec'' = Pec + S pec.  In this 
and subsequent plots, the panel labeled ``all LINERs'' represents the sum of 
pure LINERs and transition objects, and the solid histograms denote type 1 
AGNs.

Fig. 2. --- 
Distribution of morphological types for ({\it a}) all emission-line nuclei, 
\hii\ nuclei, and AGNs, and ({\it b}) the different classes of AGNs.  
In this and subsequent histograms, the downward-pointing arrow marks the 
median of the distribution.

Fig. 3. --- 
Distribution of total absolute blue magnitudes ($M_{B_T}^0$, corrected for 
internal extinction) for the host galaxies of ({\it a}) all sample objects, 
\hii\ nuclei, AGNs, and absorption-line nuclei, and ({\it b}) LINERs, 
transition objects, all LINERs, and Seyferts.  The bins are separated by 
0.5 mag.  An $L_*$ galaxy has $M_{B_T}^*\, \approx$ --20.1 mag.

Fig. 4. --- 
Distribution of equivalent widths of the narrow H\al\ emission line for 
all emission-line nuclei, \hii\ nuclei, and AGNs.  The bins are separated 
by 2 \AA, and the last bin contains all objects with EW(H\al) $>$ 30 \AA.

Fig. 5. --- 
Distribution of luminosities of the narrow H\al\ emission line for all 
emission-line nuclei, \hii\ nuclei, and AGNs.  The luminosities in the shaded 
and solid histograms were corrected for Galactic and internal reddening, 
the latter determined from the observed Balmer decrement (see Paper III), while 
the observed luminosities are shown in the unshaded histogram with heavy 
line.  The bins are separated by 0.25 in logarithmic units.

Fig. 6. --- 
Distribution of equivalent widths of the narrow H\al\ emission line as a 
function of ({\it a}) the numerical Hubble type index, T, and ({\it b}) the 
total absolute blue magnitude of the galaxy, $M_{B_T}^0$.  The numerical 
indices have the following correspondence to the Hubble sequence (see Table 12 
of Paper III): --5 = E0, --3 = S0, 1 = Sa, 3 = Sb, 5 = Sc, 7 = Sd, 10 = Im, 90 
= I0, and 99 = Pec or S pec.  The typical uncertainty in the line measurement 
is illustrated by the vertical bar in the lower right corner.

Fig. 7. --- 
Distribution of luminosities of the narrow H\al\ emission line as a function 
of ({\it a}) the numerical Hubble type index, T, and ({\it b}) the total 
absolute blue magnitude of the galaxy, $M_{B_T}^0$.  The luminosities were
corrected for Galactic and internal reddening. The typical uncertainty in 
the line measurement is illustrated by the vertical bar in the lower right 
corner.

Fig. 8. --- 
Distribution of total apparent blue magnitudes ($B_T$) for ({\it a}) all sample 
galaxies, \hii\ nuclei, and AGNs, and ({\it b}) the different classes of 
AGNs.  The bins are separated by 0.5 mag.
 
Fig. 9. --- 
Distribution of distances for ({\it a}) all sample galaxies, \hii\ nuclei, and 
AGNs, and ({\it b}) the different classes of AGNs.  The bins are separated by 
5 Mpc.

Fig. 10. --- 
Distribution of the cosine of the inclination angle for ({\it a}) all sample 
galaxies, \hii\ nuclei, and AGNs, and ({\it b}) the different classes of 
AGNs.  The bins are separated by 0.1.

%
%
%
%
%
%
%
%
%
%
%
%
%
%
%
%
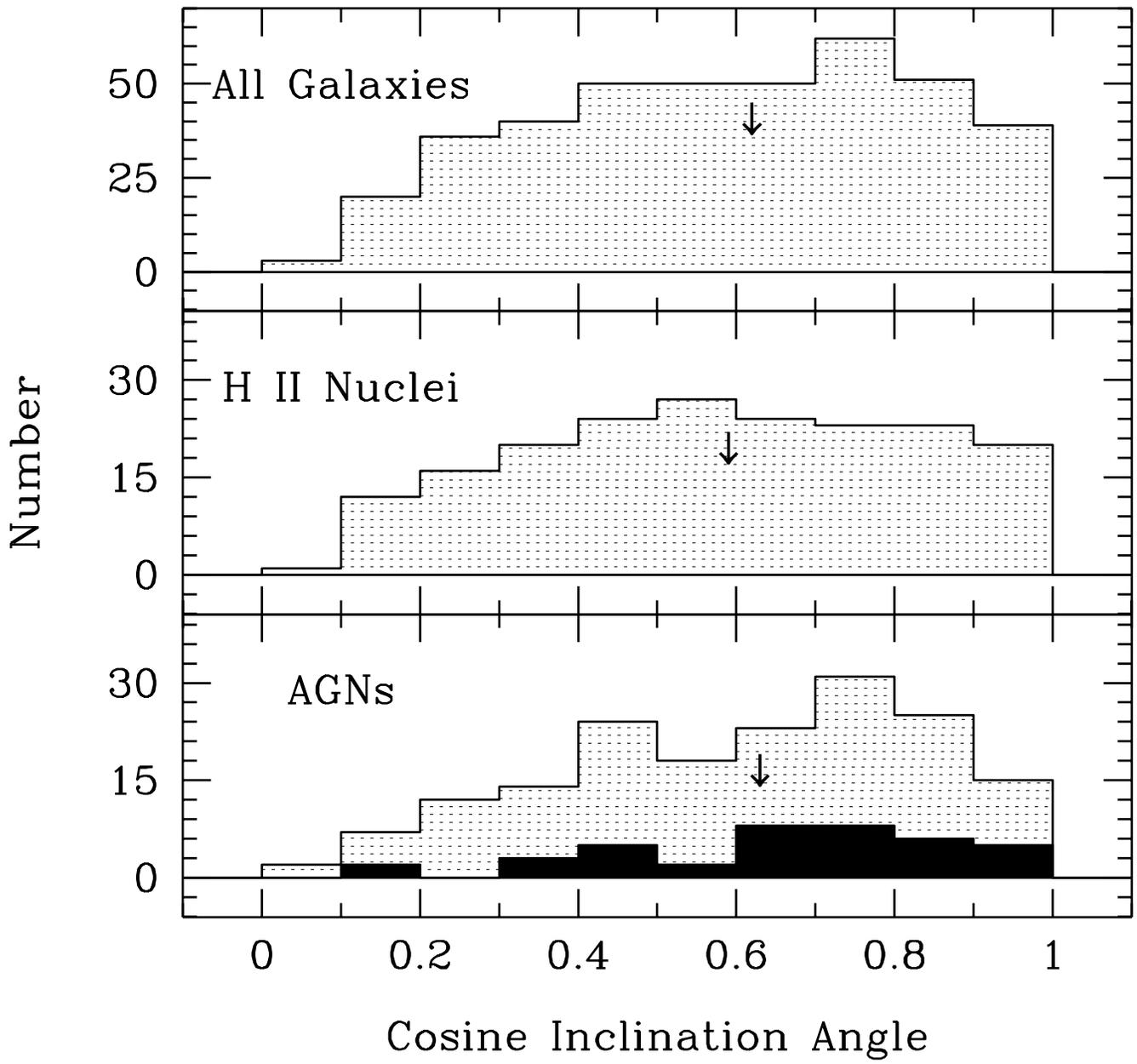
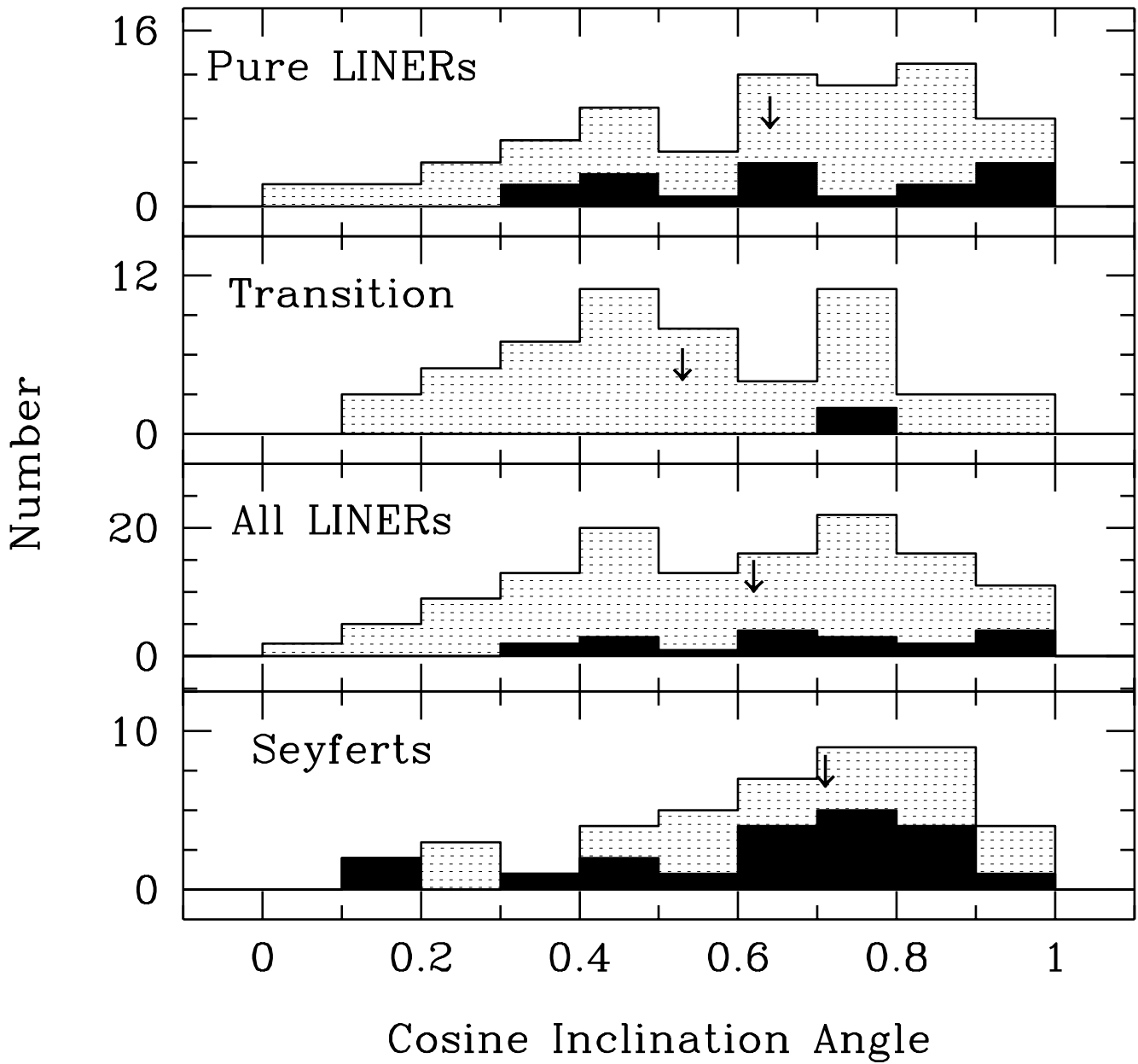
\end{document}